\documentclass[12pt]{article}
\usepackage{epsfig}
\usepackage[cp1251]{inputenc}
\usepackage[english]{babel}
\usepackage{amssymb}
\usepackage{amsmath}

\newcommand{\fr}[2]{{\displaystyle \frac{#1}{#2}}}
\newcommand{\ff}{{{f}{\!}{f}}}
\newcommand{\rf}{{{r}{\!}{f}}}
\newcommand{\bm}[1]{{\boldsymbol{\bf #1}}}
\renewcommand{\vec}[1]{{\bm{#1}}}
\newcommand{\mat}[1]{{\sf #1}}
\newcommand{\sfr}[2]{{{#1}/{#2}}}

\newcommand{\pdiff}[2]{{\fr{\partial{#1}}{\partial{#2}}}}

\newcommand{\etg}{\varepsilon_{t}}
\newcommand{\emg}{\varepsilon_{m}}
\newcommand{\ewg}{\varepsilon_{\omega}}

\title{Evolution of Collapse Nonuniformity for Rotating Magnetic Interstellar
Clouds\footnote{Submitted in Astronomy Letters, 2006, \textbf{32}, 9,
622–632. Original Russian Text: A.E. Dudorov, A.G. Zhilkin, N.Y.
Zhilkina, 2006, submitted in Pis’ma v Astronomicheskii Zhurnal, 2006,
\textbf{32}, 9, 691-702.}}
\author
{ Dudorov A.E.$^{1}$, %
  Zhilkin A.G.$^{1,2}$, %
  Zhilkina N.Y.$^{1}$\\ %
\textit{\small $^{1}$ Chelyabinsk State University, Chelyabinsk, Russia}\\%
\textit{\small $^{2}$ Institute of Astronomy RAS, Moscow, Russia, e-mail: zhilkin@inasan.ru}%
}%
\date{}

\begin{document}
\maketitle

\begin{abstract}
We investigate the formation and evolution of isothermal collapse
nonuniformity for rotating magnetic interstellar clouds. The initial
and boundary conditions correspond to the statement of the problem of
homogeneous cloud contraction from a pressure equilibrium with the
external medium. The initial uniform magnetic field is collinear with
the angular velocity. Fast and slow magnetosonic rarefaction waves
are shown to be formed and propagate from the boundary of the cloud
toward its center in the early collapse stages. The front of the fast
rarefaction wave divides the gas mass into two parts. The density,
angular velocity, and magnetic field remain uniform in the inner
region and have nonuniform profiles in the outer region. The
rarefaction wave front surface can take both prolate and oblate
shapes along the rotation axis, depending on the relationship between
the initial angular velocity and magnetic field. We derive a
criterion that separates the two regimes of rarefaction wave dynamics
with the dominant role of electromagnetic and centrifugal forces.
Based on analytical estimations and numerical calculations, we
discuss possible scenarios for the evolution of collapse
nonuniformity for rotating magnetic interstellar
clouds.
\end{abstract}


\textbf{DOI}: \textsf{10.1134/S1063773706090076}\\


\section{Introduction}

The evolution of isothermal collapse nonuniformity for interstellar
(and, in particular, protostellar) clouds is a central problem in the
theory of star formation. This problem arose immediately after the
first numerical simulations of the collapse of protostellar clouds in
the gasdynamic approximation (Bodenheimer 1968; Larson 1969; Penston
1969).

The collapse is essentially uniform under strong gravitational
nonequilibrium (Hattory et al. 1969). A major feature of the collapse
of interstellar clouds under weak gravitational nonequilibrium is its
nonuniformity (Penston 1969; Bodenheimer 1968; Larson 1969), which
becomes self-similar with time and leads to the separation of a
low-mass ($\approx 0.003 M$, where $M$ is the mass of the collapsing
cloud) opaque core and an extended envelope accreting onto it. Larson
(1969) suggested considering a rarefaction wave that is produced by a
pressure gradient at the outer boundary and that propagates through
the gas toward the cloud center with the speed of sound as the main
cause of this nonuniformity. The effect of a rarefaction wave on the
pattern of collapse was first estimated by Disney (1972). Zel’dovich
and Kazhdan (1970) investigated the dynamics of a rarefaction wave in
a self-gravitating polytropic cloud in terms of the problem of gas
outflow in to a vacuum.

The rarefaction wave generation mechanism can be easily understood in
terms of the well-known piston problem (see Landau and Lifshitz
1988). For the collapse of interstellar clouds, the contact boundary
between the cold dense cloud matter and the hot rarefied external
interstellar medium plays the role of the piston as an interface. The
gas in the inner (with respect to the contact boundary) region is
compressed under cloud self-gravity.

In the simplest case of a nonrotating cloud without any magnetic
field, the rarefaction wave front propagates through the collapsing
gas with the speed of sound. It divides the entire cloud mass into
two parts. In the inner region, the matter remains homogeneous and
collapses freely (there is no pressure gradient). In the outer
region, nonuniform profiles of density, velocity, and other
quantities are formed.

For spherically symmetric collapse of an interstellar cloud, the
rarefaction wave focusing time is defined by the dimensionless
thermal parameter $\etg=\sfr{\Pi}{E_g}$, which is the initial ratio
of the scalar pressure integral $\Pi = \int PdV$ to the magnitude of
the cloud gravitational energy $E_g$ (see Truelove et al. 1998;
Dudorov and Zhilkin 2003 (below referred to as paper 1)).

In cold clouds ($\etg \le \etg^{*} = 10/(3\pi^2) \approx 0.34$), the
rarefaction wave focusing time is $t_{*} = t_{\ff}$, where
$t_{\ff}=\sqrt{3\pi/(32G\rho_0)}$ is the free-fall time and $\rho_0$
is the initial density of the cloud. In this case, the characteristic
self-similar profiles of density $\rho \sim r^{-2}$ and velocity $v
\sim -r^{-1}$ are formed in the rarefaction wave region immediately
adjacent to the front (Larson 1969; Penston 1969; Shu 1977).
Initially, this is a narrow region, but it expands with increasing
central density. After the separation of an opaque (protostellar)
core, the gas motion near it becomes accretional with a
characteristic density profile $\rho \sim r^{-3/2}$.

In hot clouds ($\etg>\etg^{*}$), the focusing time is shorter than
the free-fall time ($t_{*}<t_{\ff}$). A nonuniform density profile is
formed in the cloud after the reflection of the rarefaction wave from
the center and a pressure gradient will affect significantly the
subsequent contraction. Since the contraction of such clouds will be
appreciably slower, this case may correspond to quasi-static
contraction of hot clouds or clouds maintained by turbulent pressure.

In rotating nonmagnetic clouds, the gas velocities along and across
the rotation axis are different due to the action of centrifugal
forces. Therefore, the surface of the rarefaction wave front becomes
oblate along the rotation axis (Tsuribe and Inutsuka 1999). In
nonrotating magnetic clouds, the magnetic field remains uniform (and,
hence, force-free) in the homogeneous region. Dudorov and Zhilkin
(paper 1) showed that the fast magnetosonic rarefaction wave (below
called the fast MHD rarefaction wave) is the main rarefaction wave
responsible for the collapse nonuniformity. Since this wave
propagates across the magnetic field lines faster than along the
magnetic field, the rarefaction wave front takes a prolate shape
along the magnetic field lines in the initial contraction stages. The
outer part of the collapsing cloud (the rarefaction wave region)
takes an oblate shape along the magnetic field lines due to the
action of electromagnetic forces.

In this paper, we investigate the dynamics of a rarefaction wave in
collapsing rotating magnetic interstellar clouds. In this case, one
might expect a great variety of rarefaction wave front shapes. To
narrow this variety and to simplify the problem, we consider the case
where the directions of the magnetic field and the angular velocity
coincide.

\section{Statement of the problem}

Let us consider a homogeneous protostellar cloud (which is a special
case of an interstellar cloud) of a given mass that is in pressure
equilibrium with the external medium.We assume that the cloud is
threaded by a uniform magnetic field $\vec{B}_0$ collinear with the
angular velocity $\vec{\Omega}_0$ at the initial time. The gas
selfgravity is initially not balanced by any forces. Therefore, a gas
motion toward the center will subsequently arise throughout the
cloud. Fast and slow MHD rarefaction waves propagating toward the
cloud center are formed at the cloud boundary as a result of
discontinuity decay (Barmin and Gogosov 1960). The fast rarefaction
wave front divides the cloud into two parts. In the inner region, the
density, angular velocity, and magnetic field remain uniform. An
inhomogeneous region is formed behind the fast rarefaction wave
front. The slow rarefaction wave propagates against the background of
this inhomogeneity and shows up as a small break in the nonuniform
profile. In this paper, we focus our attention on investigating the
dynamics of the fast rarefaction wave, which produces the
inhomogeneity and has a decisive effect on the evolution of collapse
nonuniformity.

The protostellar cloud is transparent to intrinsic infrared dust
radiation in the initial collapse stages. Therefore, we will consider
the problem of the collapse of a rotating magnetic protostellar cloud
in the approximation of ideal isothermal self-gravitational
magnetohydrodynamics. Note that this approximation must also work
well for other collapsing interstellar clouds.

The system of equations to describe self-gravitating isothermal MHD
flows can be written as
\begin{equation}\label{eq201}
  \pdiff{\rho}{t} +
  \nabla \cdot \left( \rho \vec{v} \right)
  = 0,
\end{equation}
\begin{equation}\label{eq202}
  \pdiff{\vec{v}}{t} +
  \left( \vec{v} \cdot \nabla \right) \vec{v}
  =
  -\fr{1}{\rho} \nabla P -
  \fr{1}{4\pi\rho} \left[ \vec{B}, \left[\nabla, \vec{B}\right] \right] -
  \nabla \Phi,
\end{equation}
\begin{equation}\label{eq203}
  \pdiff{\vec{B}}{t} =
  \left[\nabla, \left [ \vec{v}, \vec{B} \right]\right], \ \
  \nabla \cdot \vec{B} = 0,
\end{equation}
\begin{equation}\label{eq204}
  \nabla^2 \Phi = 4 \pi G \rho, \ \
  P = c_T^2 \rho,
\end{equation}
where $c_T$ is the isothermal speed of sound; the remaining
quantities have their universally accepted physical values.

\section{Flow configuration in the inner region}

To describe the flowof gas in a collapsing cloud, we will use the
cylindrical $(r, \varphi, z)$ coordinates. Since the problem is
axisymmetric, the variables will not depend on the azimuthal angle
$\varphi$.

In the inner region, the gas remains uniform. The magnetic field,
rotation, and the collapse itself must also remain uniform.
Therefore, the solution to Eqs. (\ref{eq201})--(\ref{eq204}) in the
inner region can be sought in the form
\begin{equation}\label{eq301}
  \rho(\vec{r}, t) = \rho(t), \ \
  \vec{B}(\vec{r}, t) = \left( 0, 0, B(t) \right),
\end{equation}
\begin{equation}\label{eq302}
  v_r(\vec{r}, t) = H_r(t) r, \ \
  v_z(\vec{r}, t) = H_z(t) z, \ \
  v_{\varphi}(\vec{r}, t) = \Omega(t) r,
\end{equation}
where $\Omega(t)$ is the angular velocity of the cloud. The radial
and vertical velocity components at each time depend linearly on the
corresponding coordinates with the proportionality coefficients
$H_r(t)$ and $H_z(t)$. A similar approach without including a
magnetic field was used by Lynden-Bell (1964) and Tsuribe and
Inutsuka (1999).

Let us change to dimensionless variables using the relations
\begin{equation}\label{eq303}
  t = t_0 \tau, \ \
  \rho(t) = \rho_0 \sigma (\tau),\ \
  B(t) = B_0 b(\tau),
\end{equation}
\begin{equation}\label{eq304}
  v_r(r, z, t) = \frac{r}{t_0} h_r(\tau), \ \
  v_z(r, z, t) = \frac{z}{t_0} h_z(\tau), \ \
  v_{\varphi}(r, z, t) = \frac{r}{t_0} \omega(\tau),
\end{equation}
where $h_r$ and $h_z$ are dimensionless analogs of the functions
$H_r$ and $H_z$, respectively. Here, the characteristic gravitational
time $t_0=\sfr{1}{\sqrt{4\pi G \rho_0}}$, the initial density
$\rho_0$, and the initial magnetic field $B_0$ are used as the main
scales.

Using the introduced dimensionless variables, we can reduce the
system of basic equations (\ref{eq201}--\ref{eq204}) to the following
system of ordinary differential equations:
\begin{equation}\label{eq306}
 \dot{\sigma} + \sigma(2h_r+h_z) = 0,
 \end{equation}
\begin{equation}\label{eq307}
  \dot{h_r} + h_r^{2} = \omega^{2} - \sigma G_r(e),
\end{equation}
\begin{equation}\label{eq308}
  \dot{h_z} + h_z^{2} = -\sigma G_z(e)
\end{equation}
\begin{equation}\label{eq309}
  \dot{\omega} + 2 h_r \omega = 0,
\end{equation}
\begin{equation}\label{eq310}
  \dot{b} + 2 h_r b = 0,
\end{equation}
where the dot denotes differentiation with respect to the
dimensionless time $\tau$.

The functions $G_r(e)$ and $G_z(e)$ define the components of the
gravitational force. Expressions for these functions can be derived
by solving the Poisson equation for the gravitational potential of a
uniform oblate ellipsoid of revolution:
\begin{equation}\label{eq312}
  G_r(e) = \fr{\sqrt{1 - e^2}}{2e^3}
  \left( \arcsin e - e \sqrt{1 - e^2} \right),
\end{equation}
\begin{equation}\label{eq313}
  G_z(e) = \fr{1}{e^3}
  \left( e - \sqrt{1 - e^2} \arcsin e \right).
\end{equation}
Here, $e$ is the eccentricity of the ellipse with the semimajor and
semiminor axes $a$ and $c$, respectively. In the solution for the
inner region, the quantities $a(\tau)$ and $c(\tau)$ act as the
spatial scales in the $r$ and $z$ directions. It is easy to verify
that they satisfy the equations
\begin{equation}\label{eq315}
 \dot{a} = a h_r, \ \
 \dot{c} = c h_z.
\end{equation}
The system of equations (\ref{eq306})--(\ref{eq315}) should be solved
with the initial conditions
\begin{equation}\label{eq316}
  \sigma(0) = b(0) = a(0) = c(0) = 1, \ \
  h_r(0) = h_z(0) = 0, \ \
  \omega(0) = \omega_0.
\end{equation}

The order of this system can be reduced significantly using the
algebraic integrals (Sedov 1981) that express the laws of
conservation of mass, angular momentum, and magnetic flux:
\begin{equation}\label{eq317}
  \sigma = \frac{1}{a^{2}c}, \ \
  \omega = \frac{\omega_0}{a^{2}},  \ \
  b = \frac{1}{a^{2}}
\end{equation}
Using Eqs. (\ref{eq315}) and (\ref{eq317}), we can reduce the system
of equations (\ref{eq306}--\ref{eq310}) to a system of two
second-order equations for the functions $a(\tau)$ and $c(\tau)$:
\begin{equation}\label{eq318}
  \ddot{a} = \fr{\ewg}{a^{3}} - \fr{G_r(e)}{a c}, \ \
  \ddot{c} = - \frac{G_z(e)}{a^{2}},
\end{equation}
where the rotational parameter $\ewg=\sfr{E_{\omega}}{E_g}$ is the
initial ratio of the rotational energy to the magnitude of the
gravitational energy of the cloud. The order of the derived system of
equations can also be reduced using the energy integral (see
Lynden-Bell 1964). However, this is not necessary, since an exact
analytical solution of system (\ref{eq318}) cannot be obtained
anyway. At the same time, it is more convenient to solve numerically
this system in form (\ref{eq318}).

It should be noted that $\ddot{a} = 0$ at $\ewg = 1/3$ at the initial
time and this value of the rotational parameter defines a centrifugal
barrier. The cloud will expand at $\ewg > 1/3$ radially. Therefore,
in our subsequent calculations, we will assume that the rotational
parameter varies within the range $0 \le \ewg \le 1/3$.

\section{Motion of the rarefaction wave front}

The $R$ coordinate of the fast MHD rarefaction wave front boundary
satisfies the equation
\begin{equation}\label{eq323}
  \frac{dR}{dt} = v(R, t) - u_f,
\end{equation}
where $v(R,t)$ is the gas flow velocity,
\begin{equation}\label{eq324}
    u_{f}
   =
    \left\{
      \frac{c_T^2 + u_A^2}{2}
     +
      \frac{1}{2}
      \left[
        (c_T^2 + u_A^2)^2
       -
        4 c_T^2 u_A^2
        \cos^2\theta
      \right]^{1/2}
    \right\}^{1/2}
\end{equation}
is the fast magnetosonic speed, $\theta$ is the angle between the
magnetic field vector ${\bf B}$ and the normal vector ${\bf n}$ to
the front surface at a given point, and $u_{A} =
\sfr{B}{\sqrt{4\pi\rho}}$ is the Alfv\`{e}n speed. We emphasize that
the velocity of the fast MHD rarefaction wave front through a
collapsing gas (\ref{eq324}) is determined only by the characteristic
structure of the MHD equations (\ref{eq201})--(\ref{eq204}).
Therefore, in general, it is not equal to the phase velocity of fast
magnetosonic waves. A more detailed justification of Eqs.
(\ref{eq323}) and (\ref{eq324}) for the velocity of the fast MHD
rarefaction wave front in a rotating magnetic cloud is given in the
Appendix.

The angle $\theta$ is 0 or $\pi$ along the magnetic field lines.
Therefore, the velocity of the rarefaction wave boundary through the
gas in the longitudinal direction is $u_{\parallel} = \max
\left\{c_T, u_A\right\}$. In the transverse direction
($\theta=\pm\pi/2$), this boundary moves through the gas with the
velocity $u_{\perp} = \sqrt{c_{T}^{2} + u_A^2}$. Let us analyze the
propagation of the fast rarefaction wave front only in the
longitudinal (along the $z$ coordinate) and transverse (along the $r$
coordinate) directions. Denoting the corresponding coordinates of the
front surface by $R_{\rf}$ and $Z_{\rf}$, we obtain
\begin{equation}\label{eq328}
  \frac{dR_{\rf}}{dt} = v_r(R_{\rf}, t) - u_{\perp}, \ \
  \frac{dZ_{\rf}}{dt} = v_z(Z_{\rf}, t) - u_{\parallel}.
\end{equation}

It should be noted that $u_{\parallel} < u_{\perp}$. However, at the
same distance from the cloud center, the radial gas velocity will be
lower than the longitudinal one due to the action of centrifugal
forces. Therefore, for a given time, the rarefaction wave front
surface in a rotating magnetic cloud can be both prolate and oblate
along the rotation axis.

Changing to the dimensionless variables $r_{\rf}=\sfr{R_{\rf}}{R_0}$
and $z_{\rf}=\sfr{Z_{\rf}}{R_0}$ in Eqs. (\ref{eq325}), where $R_0$
is the initial radius of the cloud, we transform them to
\begin{equation}\label{eq330}
  \dot{\xi}=
   - \fr{1}{a} \sqrt{\alpha_t^2 + \alpha_m^2 \fr{c}{a^{2}}}, \ \
  \dot{\zeta}=
   - \fr{1}{c}
   \max{\left(\alpha_t, \alpha_m \fr{\sqrt{c}}{a}\right)},
\end{equation}
where $\xi = \sfr{r_{\rf}}{a}$, $\zeta = \sfr{z_{\rf}}{c}$, $\alpha_t
= \sqrt{\sfr{\etg}{5}}$, $\alpha_m = \sqrt{\sfr{2\emg}{5}}$, $\emg =
\sfr{E_m}{E_g}$ is the initial ratio of the magnetic energy to the
magnitude of the gravitational energy of the cloud. Equations
(\ref{eq330}) with the initial conditions $\xi(0)=\zeta(0)=1$ must be
solved together with the system of equations (\ref{eq318}).

It should be noted that the Alfv\`{e}n speed $u_A$ in the inner
region of collapsing rotating magnetic protostellar clouds varies
with time in a more complex way than it does in the case of
nonrotating clouds. It can be easily shown that, in this case, it
initially increases, reaching a maximum at a certain time, and then
begins to decrease. The value of this maximum and the time at which
it is reached are defined by the parameters $\etg$, $\emg$ and
$\ewg$, which characterize the initial state of the cloud.

\section{Shape of the rarefaction wave front surface}

Generally, no analytical solution of the system of equations
(\ref{eq318}) and (\ref{eq330}) can be obtained. The problem can be
simplified significantly in the slow-rotation approximation where
$\ewg$ is a small parameter. In this case, the equations that
describe the rarefaction wave dynamics can be solved approximately
using a perturbation analysis. In this approximation, the values of
$a$, $c$, $r_{\rf}$ and $z_{\rf}$ can be sought in the form of an
expansion in a power series of $\ewg$. Retaining the first several
terms of the series (the order of smallness of the approximation), we
can derive equations for the coefficients of the powers of $\ewg$. We
derived explicit equations for these functions in the first
perturbation order (see Dudorov et al. 2004).

In this paper, to investigate the dynamics of the fast MHD
rarefaction wave in collapsing protostellar clouds, we numerically
solved Eqs. (\ref{eq318}) and (\ref{eq330}) using the fourth-order
Runge–Cutta method. Note that the solutions of these equations depend
on three parameters, $\etg$, $\emg$ and $\ewg$, which characterize
the initial state of the cloud. It makes sense to consider separately
a nonrotating magnetic cloud, $\ewg=0$, a rotating nonmagnetic cloud,
$\emg=0$, and a rotating magnetic cloud, $\emg \ne 0$, $\ewg \ne 0$.
In all cases, the thermal parameter $\etg$ is equal to the critical
value of $\etg^{*}$ (see the Introduction).\\

\textit{A nonrotating magnetic cloud.}\\

In the inner region of a collapsing nonrotating magnetic protostellar
cloud, the magnetic field remains uniform (and, hence, force-free)
and varies with time as $B \sim \rho^{2/3}$. Therefore, the gas
velocity in the inner region can be determined by solving the problem
of free-fall collapse. The weak discontinuity surface moves through
the gas with the fast magnetoacoustic speed that depends on the angle
between the magnetic field vector and the normal vector to a given
point of the wave front surface. In paper 1, we obtained the
analytical solutions of Eq. (\ref{eq330}) for $r_{\rf}$ and $z_{\rf}$
that correspond to this case.

\begin{figure}[t]
\begin{center}
\epsfig{file=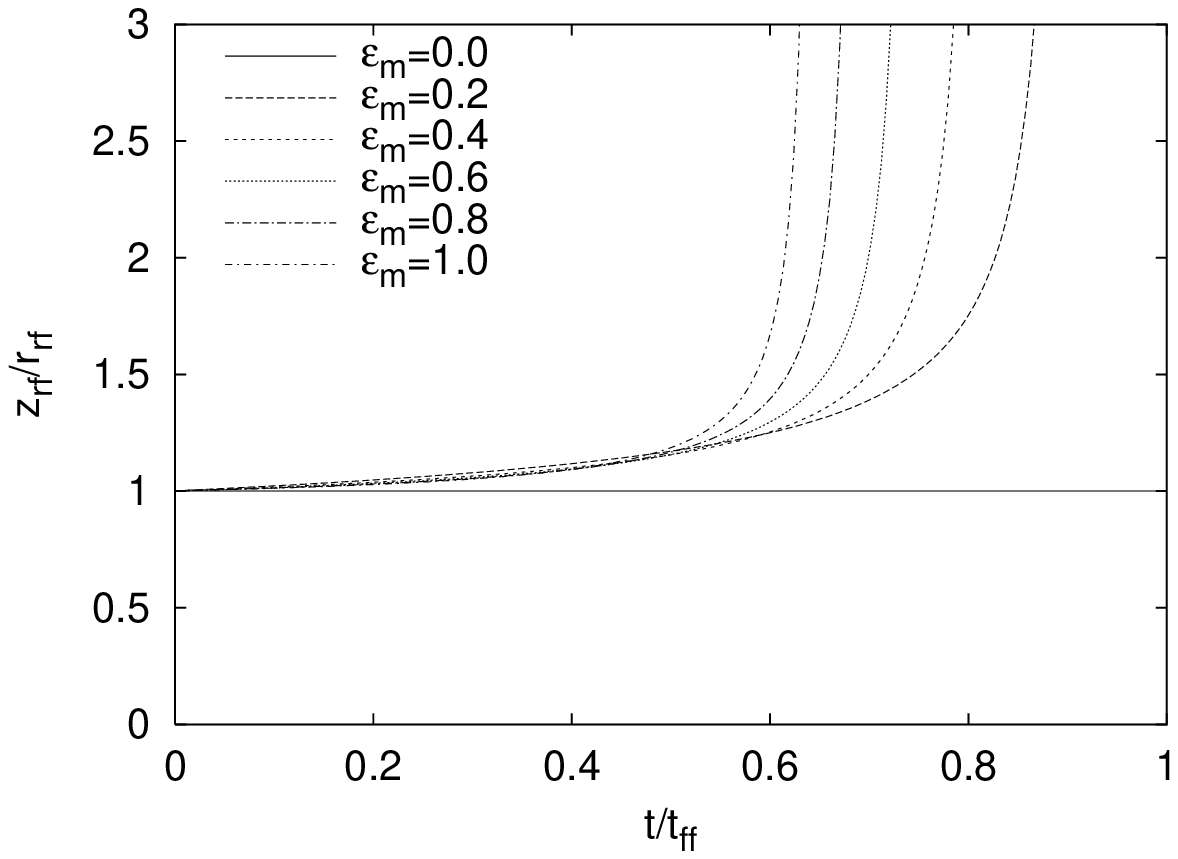,width=\textwidth}%
\caption{Degree of elongation $z_{\rf}/r_{\rf}$ of the rarefaction
wave front surface in a collapsing nonrotating magnetic cloud vs.
time. Different curves correspond to different values of the
parameter $\emg$.}
\end{center}
\label{fg1}
\end{figure}

In Fig. \ref{fg1}, the ratio $z_{\rf}/r_{\rf}$, which defines the
degree of elongation of the rarefaction wave front surface along the
magnetic field, is plotted against time. Different curves in the
figure correspond to different values of the parameter $\emg$, which
characterizes the initial magnetic field. The figure shows that,
while $r_{\rf}$ and $z_{\rf}$ generally decrease, their ratio
$z_{\rf}/r_{\rf}>1$ and infinitely increases in a finite time.
Consequently, the shape of the rarefaction wave front surface in a
collapsing nonrotating magnetic cloud is prolate along the magnetic
field lines. The rarefaction wave focusing time defines the end of
the initial cloud contraction stage. At this time, the homogeneous
region disappears and the cloud subsequently evolves against the
background of nonuniform contraction.\\

\textit{A rotating nonmagnetic cloud.}\\

In a collapsing rotating nonmagnetic protostellar cloud, the weak
discontinuity surface moves through the gas with the speed of sound
$c_T$. The gas velocity along the rotation axis is higher than that
in the transverse direction due to the action of centrifugal forces.
Therefore, the shape of the rarefaction wave front surface in this
case must be oblate along the rotation axis (see Tsuribe and Inutsuka
1999).

\begin{figure}[t]
\begin{center}
\epsfig{file=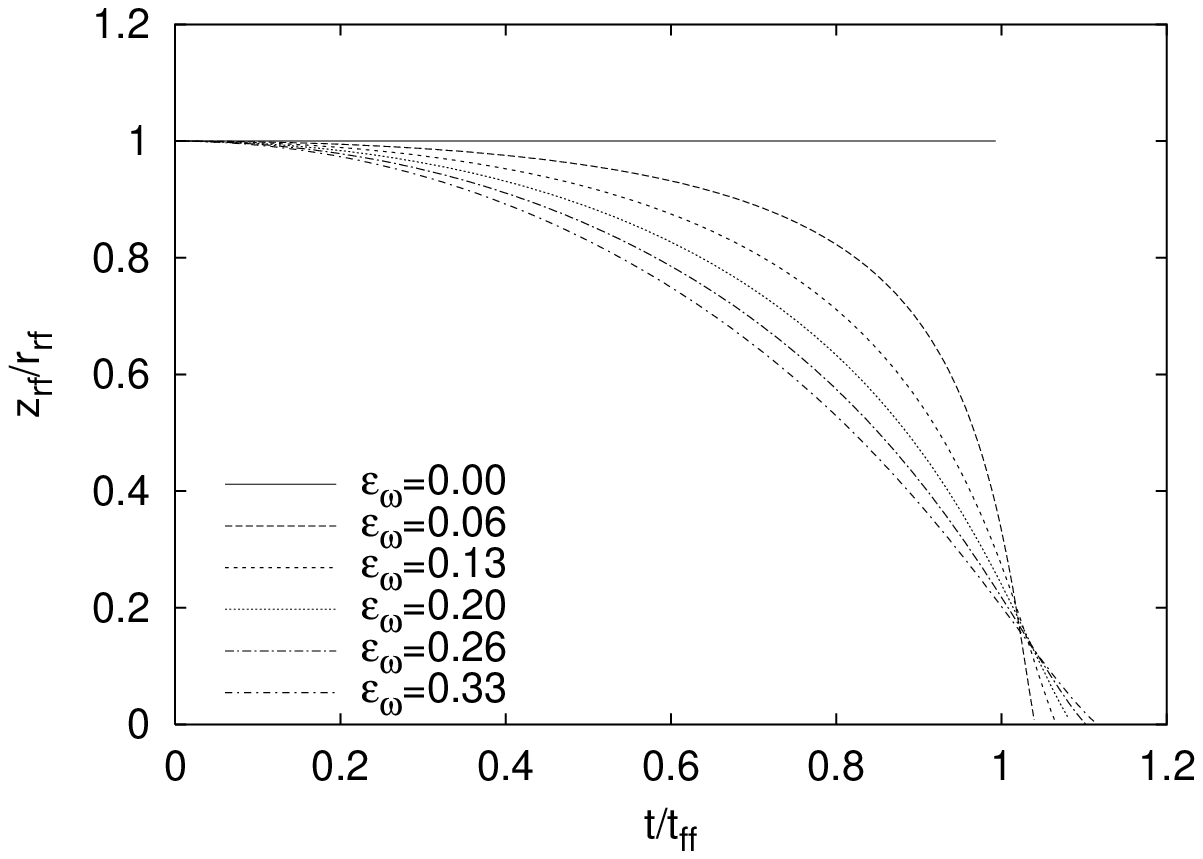,width=\textwidth}%
\caption{Degree of flattening $z_{\rf}/r_{\rf}$ of the rarefaction
wave front surface in a collapsing rotating nonmagnetic protostellar
cloud vs. time. Different curves correspond to different values of
the parameter $\ewg$.}
\end{center}
\label{fg2}
\end{figure}

In Fig. \ref{fg2}, the ratio $z_{\rf}/r_{\rf}$ is plotted against
time. In this case, it defines the degree of flattening of the
rarefaction wave front surface. Different curves in the figure
correspond to different values of the parameter $\ewg$, which
characterizes the initial rotation of the cloud. The figure shows
that, while $r_{\rf}$ and $z_{\rf}$ generally decrease, their ratio
$z_{\rf}/r_{\rf} < 1$ and decreases to zero in a finite time.\\

\textit{A rotating magnetic cloud.}\\

\begin{figure}[t]
\begin{center}
\epsfig{file=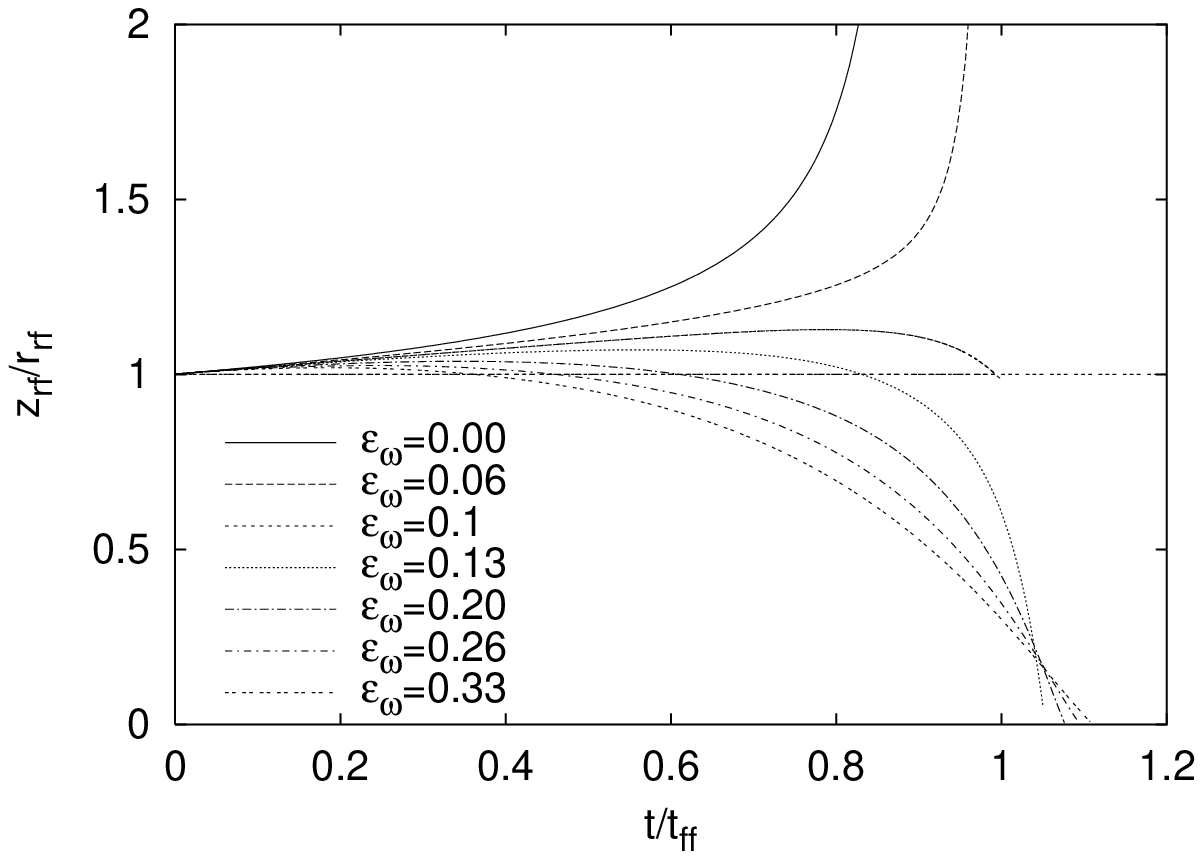,width=\textwidth}%
\caption{Degree of flattening/elongation $z_{\rf}/r_{\rf}$ of the
rarefaction wave front surface in a collapsing rotating magnetic
protostellar cloud vs. time. The parameter $\emg = 0.2$. Different
curves correspond to different values of the parameter $\ewg$.}
\end{center}
\label{fg3}
\end{figure}

In a rotating magnetic cloud, both mechanisms considered above are in
operation. Therefore, the shape of the rarefaction wave front surface
in such clouds can evolve in a complex way. In Fig. \ref{fg3}, the
ratio $z_{\rf}/r_{\rf}$ for rotating magnetic clouds is plotted
against time for $\emg = 0.2$. Different curves in the figure
correspond to different values of the parameter $\ewg$.

Analysis of the behavior of the curves in the figure leads us to
conclude that both rarefaction wave evolution scenarios are possible
in collapsing rotating magnetic protostellar clouds. If the rotation
is slow, then the magnetic field has a stronger effect on the
rarefaction wave dynamics and the wave front surface takes a prolate
shape along the magnetic field lines. In this case, $z_{\rf}/r_{\rf}$
infinitely increases with time (the two upper curves in Fig.
\ref{fg3}). In the case of fast rotation, the centrifugal force is
dominant. Therefore, the shape of the rarefaction wave front surface
becomes oblate along the rotation axis with time, while the ratio
$z_{\rf}/r_{\rf}$ decreases to zero with time (the lower curves in
Fig. \ref{fg3}).

Interestingly, the shape of the rarefaction wave front surface in
collapsing rotating magnetic protostellar clouds is always prolate
along the rotation axis in the initial stage. In Fig. \ref{fg3}, all
curves initially run above the straight line $z_{\rf}/r_{\rf} = 1$
and only after a lapse of time does the ratio $z_{\rf}/r_{\rf}$
becomes smaller than unity in the case of fast rotation.

The critical case where the effects of magnetic field and rotation on
the rarefaction wave dynamics are balanced separates the two
described rarefaction wave evolution scenarios in collapsing rotating
magnetic protostellar clouds. Therefore, the rarefaction wave is
focused in the longitudinal and transverse directions almost
simultaneously. In Fig. \ref{fg3}, $\ewg \approx 0.1$ corresponds to
this case.

\section{The focusing time}

The focusing time $t_{*}$ is defined as the time in which the
rarefaction wave front surface reaches the cloud center. The focusing
time depends on three parameters: $\etg$, $\emg$ and $\ewg$.

\begin{figure}[t]
\begin{center}
\epsfig{file=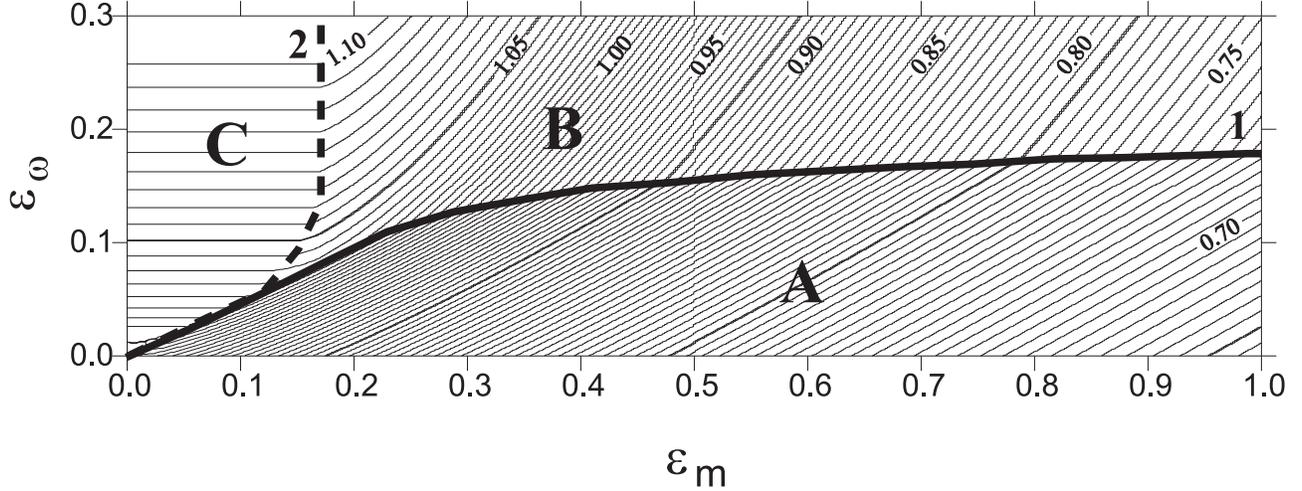,width=\textwidth}%
\caption{Lines of equal focusing time in the $\emg$, $\ewg$ plane
calculated for $\etg=\etg^{*}$. Solid line 1 corresponds to the
critical curve that separates the two rarefaction wave evolution
scenarios with the dominant role of rotation (regions B and C) and
magnetic field (region A). Dashed curve 2 highlights region C where
the focusing time does not depend on the magnetic parameter $\emg$.}
\end{center}
\label{fg4}
\end{figure}

Figure \ref{fg4} shows the curves of equal focusing time $t_{*}(\emg,
\ewg) = \text{const}$ in the $\emg$, $\ewg$ plane in the case where
the thermal parameter $\etg=\etg^{*}$. The numbers on the curves
indicate the focusing times calculated in units of the free-fall time
$t_{\ff}$. This figure shows that the focusing time decreases with
increasing magnetic parameter $\emg$ (at fixed $\ewg$) and increases
with increasing rotational parameter $\ewg$ (at fixed $\emg$). This
pattern of the dependence $t_{*}(\emg, \ewg)$ can be easily
explained. The fast magnetosonic speed increases with growing
magnetic field; therefore, the focusing time $t_{*}$ must decrease
with increasing magnetic parameter $\emg$. On the other hand, the
centrifugal force increases with increasing angular velocity and,
hence, the velocity of the collapsing gas slows down. Therefore, the
focusing time $t_{*}$ must increase with increasing rotational
parameter $\ewg$.

The curves of equal focusing time undergo a break on the two lines
denoted by 1 and 2. Heavy solid line 1 corresponds to the critical
case where the effects of electromagnetic and centrifugal forces on
the dynamics of the fast MHD rarefaction wave are balanced near the
focusing time. The front surface near the focusing time has a nearly
spherical shape ($z_{\rf}$ and $r_{\rf}$ tend to zero simultaneously
as $t \to t_{*}$). This critical curve separates two regions of
parameters $\ewg$ and $\emg$. In region A below the critical curve,
the magnetic field has a stronger effect on the rarefaction wave
dynamics (and on the collapse as a whole). In this case, the
rarefaction wave front surface has a prolate shape along the rotation
axis and, hence, the focusing is transverse. In regions B and C above
the critical curve, the rarefaction wave evolves with the dominant
role of rotation. Near the focusing time, the rarefaction wave front
surface has an oblate shape along the rotation axis. The relationship
between the rotational, $\ewg$, and magnetic, $\emg$, parameters on
critical curve 1 can be roughly described by the empirical relation
\begin{equation}\label{eq601}
  \ewg = \fr{\emg}{2 + q \emg^{3/2}}.
\end{equation}
The parameter $q$ depends on $\etg$. For $\etg=\etg^{*}$, $q = 3.4$.

In regions B and C, the rarefaction wave is focused in the
longitudinal direction. In this case, the fast magnetosonic speed is
$u_{\parallel} = \max \left\{c_T, u_A\right\}$. In region C (a weak
magnetic field), the Alfv\`{e}n speed $u_A$ is lower than the
isothermal speed of sound $c_T$. Therefore, the focusing time $t_{*}$
in this region does not depend on the magnetic parameter $\emg$ and
is the same as that for a rotating nonmagnetic cloud. The
relationship between $\ewg$ and $\emg$ on curve 2 that separates
regions B and C can be found analytically using a perturbation
analysis in the slow-rotation approximation (see Dudorov et al.
2004):
\begin{equation}
  \emg =
  \left \{
   \begin{array}{cc}
    6\etg\ewg \fr{4-15\ewg}{5-12\ewg} & \ewg \le \sfr{1}{6} \\
    \sfr{\etg}{2} & \ewg > \sfr{1}{6}
   \end{array}
  \right.
\end{equation}

The focusing time also depends on the thermal parameter $\etg$. We
analyzed the behavior of the critical curves that separate regions A,
B, and C as a function of the thermal parameter. As $\etg$ increases,
curve 1 shifts upward, while curve 2 shifts rightward. Thus, the size
of region A in which the rarefaction wave evolves with the dominant
role of magnetic field increases with increasing $\etg$. Accordingly,
the size of the region in which the rarefaction wave evolves with the
dominant role of rotation decreases. The relative size of region C,
in which the focusing time does not depend on the magnetic parameter
$\emg$, also increases.

\section{Comparison of analytical solutions with numerical simulations}

The results of our analysis of the fast MHD rarefaction wave dynamics
are in good agreement with our direct numerical simulations of the
collapse of rotating magnetic protostellar clouds in the 2D
approximation. The computations were performed on a $300 \times 800$
grid in Euler variables in cylindrical coordinates using a numerical
MHD code (Dudorov et al. 1999a) that is based on the total variation
diminishing (TVD) scheme for MHD equations (Dudorov et al. 1999b).

\begin{figure}[t]
\begin{center}
\epsfig{file=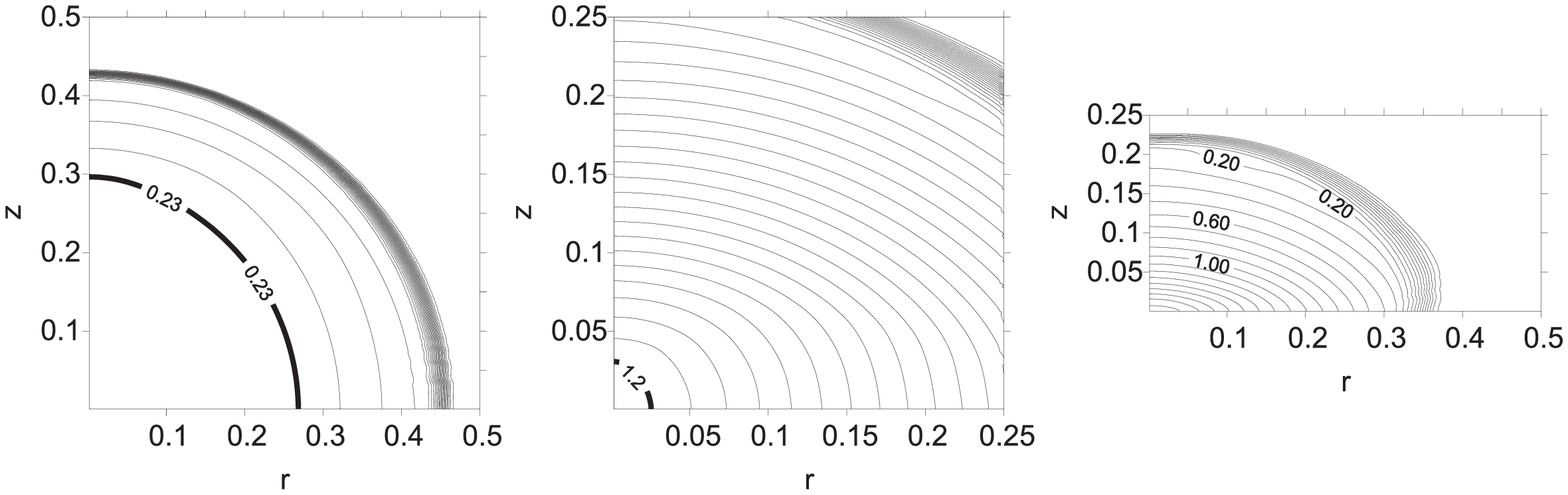,width=\textwidth}%
\caption{Density logarithm distribution and shape of the fast MHD
rarefaction wave front surface (heavy lines) in a collapsing rotating
magnetic protostellar cloud with the initial parameters
$\etg=\etg^{*}$, $\emg=0.2$, and $\ewg=0.05$. The times $0.53
t_{\ff}$, $0.93 t_{\ff}$, and $1.03 t_{\ff}$, respectively, are shown
from left to right.}
\end{center}
\label{fg5}
\end{figure}

\begin{figure}[t]
\begin{center}
\epsfig{file=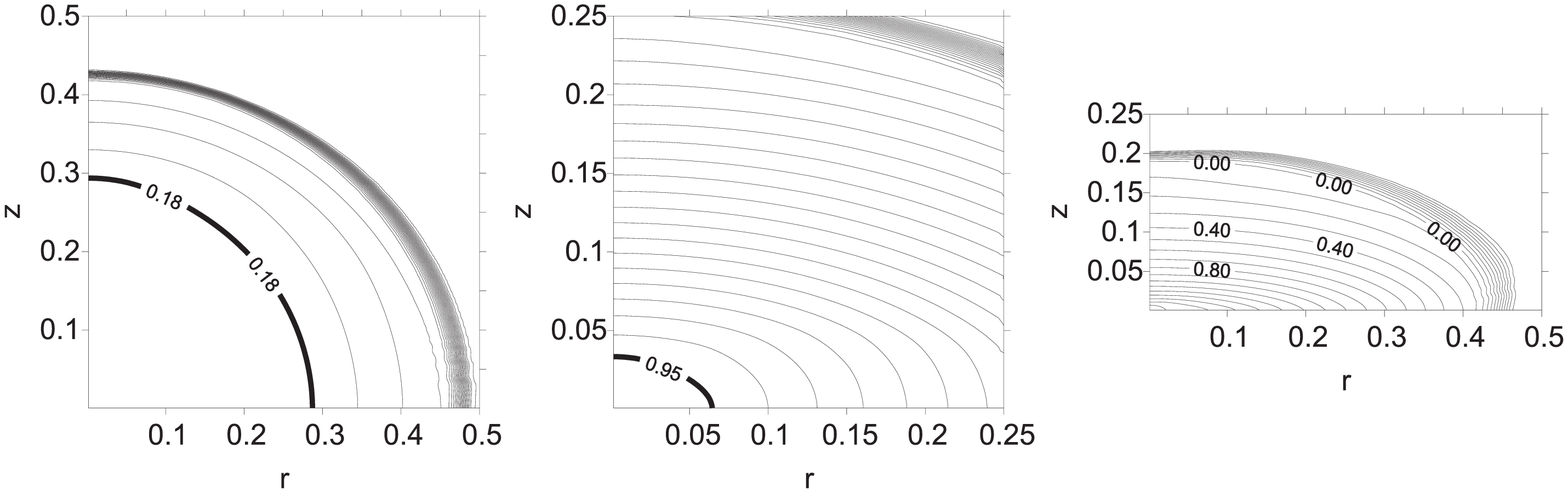,width=\textwidth}%
\caption{Density logarithm distribution and shape of the fast MHD
rarefaction wave front surface (heavy lines) in a collapsing rotating
magnetic protostellar cloud with the initial parameters
$\etg=\etg^{*}$, $\emg=0.2$, and $\ewg=0.15$. The times $0.54
t_{\ff}$, $0.98 t_{\ff}$, and $1.09 t_{\ff}$, respectively, are shown
from left to right.}
\end{center}
\label{fg6}
\end{figure}

Figures \ref{fg5} and \ref{fg6} present two cases of our numerical
simulation of the collapse of rotating magnetic protostellar clouds.
In both cases, the initial parameters of the clouds correspond to the
thermal and magnetic parameters $\etg=\etg^{*}=0.34$ and $\emg=0.2$,
respectively. The rotational parameter is $\ewg=0.05$ in the first
case (Fig. \ref{fg5}) and $\ewg=0.15$ in the second case (Fig.
\ref{fg6}). The rotational parameter in the first case was chosen in
such a way that the initial state of the cloud satisfied the
conditions of region A (see Fig. \ref{fg4}), in which the dynamics of
the fast MHD rarefaction wave is dominated by electromagnetic forces.
In the second case of our simulation, the chosen initial model
parameters satisfied the conditions of region B, in which the
dynamics of the fast MHD rarefaction wave is dominated by centrifugal
forces.

Figure \ref{fg5} (the left and middle panels) shows the density
distributions and positions of the fast MHD rarefaction wave front
(heavy solid line) for the first case of our simulation for two
times, $0.53 t_{\ff}$ and $0.93 t_{\ff}$. The numbers on the isolines
indicate the density logarithms. The rarefaction wave surface takes a
prolate shape along the rotation axis similar to the shape of a
prolate ellipsoid of revolution. In this case, the velocity of the
collapsing gas slows down in the radial direction in the
inhomogeneous region behind the rarefaction wave front due to the
action of electromagnetic and centrifugal forces. Therefore, the
cloud takes a flattened shape in the course of time. The middle panel
corresponds to a time close to the focusing time. The shape of the
rarefaction wave surface becomes highly prolate by this time. The
right panel shows the density distribution in the cloud at time $1.03
t_{\ff}$ after the rarefaction wave focusing. The homogeneous region
disappears by this time and the subsequent collapse proceeds against
the background of a nonuniform density profile. In this case of our
simulation, a cloud with a flattened disklike structure is formed in
the final contraction stages.

The results of our numerical simulation in the second case are
presented in Fig. \ref{fg6}. As in the first case, the left and
middle panels show the density distribution and the positions of the
fast MHD rarefaction wave front (heavy solid line) for times $0.54
t_{\ff}$ and $0.98 t_{\ff}$. The rarefaction wave surface takes an
oblate shape along the rotation axis similar to the shape of an
oblate ellipsoid of revolution by the time $0.98 t_{\ff}$. An oblate
shape of the cloud is also formed behind the rarefaction wave front
in the inhomogeneous region. In contrast to the previous case of our
simulation, the rarefaction wave surface is identical in shape to the
cloud configuration forming in the final contraction stage (see the
right panel).

The focusing time $t_{*}$ of the fast MHD rarefaction wave determined
by our numerical simulations is $0.97 t_{\ff}$ for the first case and
$1.06 t_{\ff}$ for the second case. These values closely match the
focusing times calculated analytically in this section (see Fig.
\ref{fg4}).

\section{Discussion and conclusions}

In this paper, we have considered the formation of collapse
nonuniformity for rotating magnetic protostellar clouds. Note that
all of our results formulated for protostellar clouds are also valid
for isothermal interstellar clouds. Within the framework of our
statement of the problem of homogeneous cloud contraction in a
pressure equilibrium with the external medium, the collapse dynamics
is characterized by the generation of fast and slow MHD rarefaction
waves at the cloud boundary and their subsequent propagation toward
the cloud center. The surface of the fast MHD rarefaction front
divides the entire volume of the collapsing cloud into two regions.
In the inner region, the gas remains homogeneous and is characterized
by uniform rotation and magnetic field. In this region, the pressure
gradient is zero. In the outer region, nonuniform density, velocity,
magnetic field, and angular velocity profiles are formed. The degree
of nonuniformity can increase greatly with time. The slow MHD
rarefaction wave propagates in the wake of the fast one against the
background of an evolving nonuniformity, acting as a generator of
perturbations in this region. Thus, the fast MHD rarefaction wave is
mainly responsible for the collapse nonuniformity of rotating
magnetic protostellar clouds. Its parameters (the front velocity and
surface shape) determine the rate of evolution and degree of
inhomogeneity of collapsing clouds.

Depending on the relationship between the parameters that
characterize the initial magnetic field and rotation of the cloud,
the shape of the fast MHD rarefaction wave surface can be both
prolate and oblate along the rotation axis. Analyzing the rarefaction
wave dynamics, we can identify two scenarios for the collapse of
rotating magnetic protostellar clouds.

In the first case, the collapse takes place with the dominant role of
a magnetic field. The surface of the fast MHD rarefaction wave has a
prolate shape along the rotation axis and it is focused in the
direction transverse to the magnetic field. In the second case, the
collapse takes place with the dominant role of rotation. The surface
of the fast MHD rarefaction wave has an oblate shape along the
rotation axis and it is focused across the magnetic field. Here, we
derived a criterion separating these two regimes of collapse (see
(\ref{eq601})).

The cores of interstellar molecular clouds may be considered to be
observational manifestations of protostellar clouds (Dudorov 1991). A
direct observational confirmation of the gravitational collapse of
some molecular cloud cores is the presence of characteristic
signatures of contraction in molecular spectra (Tafalla et al. 1998;
Williams et al. 1999; Gregersen and Evans 2000). The density
distribution in the central parts of protostellar clouds is
essentially uniform (Beuther et al. 2002; Caselli et al. 2002). For
some clouds (L1536, L1512, L1498, L1544, L1495, TMC- 2, and others),
there is observational evidence for the presence of weak
discontinuities that separate the inner homogeneous region from the
outer inhomogeneous region (Caselli et al. 2002). This may be
considered as evidence for the existence of rarefaction waves
propagating in these clouds. The shapes of the clouds themselves are
also in satisfactory agreement with theoretical predictions. For
example, an inner compact core of a nearly spherical or prolate
(along the symmetry axis) shape and an extended oblate envelope are
clearly identified in the clouds L1495, L1527, TMC-2, Per 5, Per 7,
and L1582A. In other clouds (e.g., Per 6, L1400K, TMC-1, L260, and
L1221), the central quasi-homogeneous core has a distinctly flattened
shape along the symmetry axis. The clouds L1512 and L234A have double
cores against the background of an oblate inhomogeneous envelope
along the symmetry axis. This is probably because these clouds are
gravitationally fragmented due to their rapid rotation. It should be
noted that the observed density profiles in protostellar clouds given
in the papers cited above were averaged over all directions. From the
viewpoint of this paper, it would be interesting to compare the
density profiles in the longitudinal and transverse directions with
respect to the symmetry axis of these clouds, which is defined by the
directions of the angular velocity and the largescale magnetic field.

In the outer inhomogeneous region (behind the fastMHDrarefaction wave
front), differential rotation must lead to intense generation of a
toroidal magnetic field. The toroidal magnetic field produces a
braking torque that contributes to the redistribution of angular
momentum between the central parts of the protostellar cloud and its
periphery. Depending on the relationship between the parameters
$\emg$ and $\ewg$, the magnetic braking of the cloud rotation can be
effective or ineffective (Dudorov et al. 2004). Therefore, combining
this criterion with the criterion associated with the rarefaction
wave gives four fundamentally differentMHD collapse scenarios. A
detailed analysis of these scenarios is the subject of a special
paper. However, it is worth noting that in the case of ineffective
magnetic braking, the angular momentum can be lost through other
mechanisms (fragmentation, jet outflows, etc.). It should be
emphasized once again that all these effects in protostellar clouds
arise from the collapse nonuniformity produced by MHD rarefaction
waves.

In all cases, the fast MHD rarefaction wave front surface in
collapsing rotating magnetic protostellar clouds is nonspherical in
shape. Therefore, its focusing and subsequent reflection from the
center can be accompanied by the generation of (also nonspherical)
intense nonlinear MHD waves that must affect the subsequent collapse
dynamics. If the focusing time $t_{*}$ is close to the free-fall time
$t_{\ff}$ (weak magnetic field, slow rotation, $\etg\le\etg^{*}$),
then the focusing will be accompanied by adiabatic gas heating in the
central part of the cloud. The rise in gas temperature increases the
rarefaction wave front velocity and the focusing can occur before an
infinite density is reached at the cloud center (Zel’dovich and
Kazhdan 1970). If the focusing time is shorter than the free-fall
time (strong magnetic field, rapid rotation, $\etg > \etg^{*}$), then
the focusing can occur even at the stage of isothermal contraction.
Thus, in all cases, the rarefaction wave focusing acts as a physical
factor that limits the density growth during collapse.

Thus, the fast MHD rarefaction wave that emerges in the early
contraction stages not only allows the collapse nonuniformity for
interstellar clouds to be explained, but also is a good tool for
studying this astrophysical phenomenon. It should be noted that the
conclusions reached here using semi-analytical methods are in good
agreement with the direct numerical simulations of the collapse of
rotating magnetic protostellar clouds that we have performed over
several years in the 1.5-D, 2-D, and 3-D approximations.\\

\textbf{Acknowledgement.} This work was supported by the Russian
Foundation for Basic Research (project nos. 05-02-17070, 05-02-16123,
and 04-02-96050 RFBR–Ural).

\section*{Appendix}

Let us derive an expression for the velocity of a rarefaction wave
front in a rotating magnetic cloud. The rarefaction wave front is the
surface of a weak discontinuity on which all MHD quantities remain
continuous, while their derivatives undergo a discontinuity. MHD
equations (\ref{eq201})--(\ref{eq204}) should be used in integral
form to describe MHD flows with weak discontinuities (see Kulikovskii
et al. 2001):
\begin{equation}\label{eqa1}
  \pdiff{}{t}
  \int\limits_{V}
  \bm{u}dV +
  \sum\limits_{k=1}^{3}
  \oint\limits_{\partial V}
  \bm{F}_{k} dS_{k} =
  \int\limits_{V}
  \bm{R}dV,
\end{equation}
where $\bm{u}$ is the vector of conservative variables,
$\bm{F}_k=\left(\bm{F}_x,\bm{F}_y,\bm{F}_z\right)$ are the flux
vectors in the $x$, $y$ and $z$ directions in Cartesian coordinates,
and $\bm{R}$ is the source vector, which can include, for example,
the gravitational force, the centrifugal force, the Coriolis force,
and the like. We do not write out explicit expressions for these
vectors to save space. The integration in (\ref{eqa1}) is over a
certain stationary volume $V$ bounded by the surface $\partial V$ and
$dS_{k}$ is an oriented element of this surface.

Consider a certain surface of a strong MHD discontinuity (see Fig.
\ref{fg07}). Let us choose a small portion of this surface and
construct a normal vector $\vec{n}=\left(n_x, n_y, n_z\right)$ on it.
As the volume $V$, we choose a cylinder with height $h$ and base area
$S$. The discontinuity surface divides this cylinder into two
regions, $V_R$ and $V_L$ (above and below the surface in the figure).

\begin{figure}[t]
\begin{center}
\epsfig{file=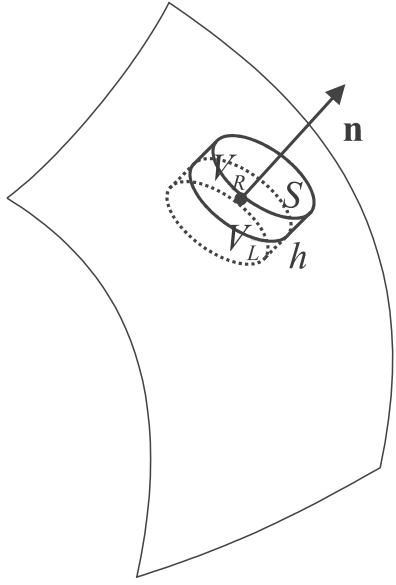,height=8cm}%
\caption{To the derivation of Hugoniot conditions on the surface of
an MHD discontinuity.}
\end{center}
\label{fg7}
\end{figure}

The limiting quantities to the left, $\bm{u}_L$, and to the right,
$\bm{u}_R$, of the discontinuity surface near the selected small
portion are related by the Hugoniot conditions. To derive these
conditions, we will shrink the cylinder to the discontinuity surface
($h \to 0$) while leaving the base areas fixed. Simple calculations
using (\ref{eqa1}) yield
\begin{equation}\label{eqa2}
  -D_n[\bm{u}]S +
  \int\limits_{V_R} \pdiff{\bm{u}}{t}dV +
  \int\limits_{V_L} \pdiff{\bm{u}}{t}dV +
  \sum\limits_{k=1}^{3}
  n_k [\bm{F}_k] S +
  \sum\limits_{k=1}^{3}
  \int\limits_{S_1} \bm{F}_{k} dS_{k} =
  \int\limits_{V}
  \bm{R}dV,
\end{equation}
where $D_n$ is the velocity of the discontinuity surface along the
normal vector $\vec{n}$ and the square brackets denote the difference
between the limiting quantities $[\bm{u}]=\bm{u}_R-\bm{u}_L$. In the
limit $h\to 0$, the second and third terms on the left-hand side, the
integral over the side surface $S_1$ of the cylinder, and the
integral on the right-hand side of Eq. (\ref{eqa2}) tend to zero. As
a result, we obtain the following Hugoniot conditions on the
discontinuity surface:
\begin{equation}\label{eqa3}
  D_n[\bm{u}] = [\bm{F}_n],
\end{equation}
where $\bm{F}_n = n_x\bm{F}_x+n_y\bm{F}_y+n_z\bm{F}_z$. Note, in
particular, that the source terms $\bm{R}$ do not appear in this
relation.

To pass to the case of a weak discontinuity, we will assume that the
right limiting values of $\bm{u}_R$ differ from the left limiting
values of $\bm{u}_L=\bm{u}_0$ by infinitesimals:
$\bm{u}_R=\bm{u}_0+\delta\bm{u}$. Expanding the right-hand side of
Eq. (\ref{eqa3}) to linear terms in $\delta\bm{u}$ yields
\begin{equation}\label{eqa4}
  D_n\delta\bm{u} = \mat{A} \cdot \delta\bm{u},
\end{equation}
where $\mat{A} = \left. \partial \bm{F}_n / \partial \bm{u}
\right|_{\bm{u}=\bm{u}_0}$ is the hyperbolicity matrix of the MHD
equations. It follows from Eq. (\ref{eqa4}) that the velocity $D_n$
of a weak discontinuity coincides with one of the eigenvalues
$\lambda_{\alpha}$ of the matrix $\mat{A}$, while $\delta\bm{u}$
coincides with one of its right eigenvectors. This determines the
possible types of MHD weak discontinuities. For example, the fast MHD
weak discontinuity (the fast MHD rarefaction wave front) considered
in our paper corresponds to the eigenvalue $\lambda_{-f} = v_n-u_f$,
where $v_n$ is the normal (to the discontinuity surface) gas velocity
and $u_f$ is the fast magnetosonic speed. Note that a similar result
for the rarefaction wave front velocity can also be obtained more
formally, by considering the conditions for the derivatives of vector
u on the weak discontinuity surface.

\small
\section*{References}

\begin{enumerate}

\item A.A. Barmin and V.V. Gogosov, Dokl. Akad. Nauk SSSR
\textbf{134}, 1041 (1960) [Sov. Phys. Dokl. \textbf{5}, 961 (1960)].

\item H. Beuther, P. Schlike, K.M. Menten, et al., Astrophys. J.
\textbf{566}, 945 (2002).

\item P. Bodenheimer, Astrophys. J. \textbf{153}, 483 (1968).

\item P. Caselli, P.J. Benson, P.C. Myers, and M. Tafalla, Astrophys.
J. \textbf{572}, 238 (2002).

\item M.J. Disney, Mon. Not. R. Astron. Soc. \textbf{175}, 323
(1976).

\item A.E. Dudorov, Astron. Zh. \textbf{68}, 695 (1991) [Sov. Astron.
\textbf{35}, 342 (1991)].

\item A.E. Dudorov, A.G. Zhilkin, and O.A. Kuznetsov, Mat. Model.
\textbf{101}, 109 (1999a).

\item A.E. Dudorov, A.G. Zhilkin, and O.A. Kuznetsov, Mat. Model.
\textbf{101}, 101 (1999b).

\item A.E. Dudorov and A.G. Zhilkin, Zh. \'{E}ksp. Teor. Fiz.
\textbf{123}, 195 (2003) [JETP \textbf{96}, 165 (2003)].

\item A.E. Dudorov, A.G. Zhilkin, N.Y. Zhilkina, \textit{MHD
Rarefaction Wave as the Cause of Collapse Nonuniformity for Rotating
Magnetic Protostellar Clouds}. Proceedings of International
Scientific Conference ''VII Zababakhin Scientific Talks'', Snezhinsk,
2004, pp. 1–16, \newline
http://www.vniitf.ru/rig/konfer/7zst/reports/s3/s-3.html.

\item E.M. Gregersen and N.J. Evans, Astrophys. J. \textbf{538}, 260
(2000).

\item T. Hattory, T. Nakano, and C. Hayashi, Prog. Theor. Phys.
\textbf{42}, 4 (1969).

\item A.G. Kulikovsky, N. V. Pogorelov, and A. Yu. Semenov,
\textit{Mathematical Problems of the Numerical Solution of Hyperbolic
Systems of Equations} (Fizmatlit, Moscow, 2001) [in Russian].

\item L.D. Landau and E.M. Lifshitz, \textit{Fluid Mechanics} (Nauka,
Moscow, 1988; Pergamon Press, Oxford, 1987).

\item R.B. Larson, Mon. Not. R. Astron. Soc. \textit{145}, 271
(1969).

\item R.B. Larson, Mon. Not. R. Astron. Soc. \textit{156}, 437
(1972).

\item D. Lynden-Bell, Astrophys. J. \textit{139}, 1195 (1964).

\item M.V. Penston, Mon. Not. R. Astron. Soc. \textit{144}, 425
(1969).

\item L.I. Sedov, \textit{Similarity and Dimensional Methods in
Mechanics} (Nauka, Moscow, 1981; Academic, NewYork, 1959).

\item F.H. Shu, Astrophys. J. \textbf{214}, 488 (1977).

\item M. Tafalla, D. Mardones, P.C. Myers, et al., Astrophys. J.
\textbf{504}, 900 (1998).

\item K. Truelove, R. I. Klein,C. F.McKee, et al., Astrophys. J.
\textbf{495}, 821 (1998).

\item T. Tsuribe and S. Inutsuka, Astrophys. J. \textbf{526}, 307
(1999).

\item J.P. Williams, P.C. Myers, D.J. Wilner, and J. Di Francesco,
Astrophys. J. \textbf{513}, L61 (1999).

\item Ya. B. Zel’dovich and Ya. M. Kazhdan, Astrofizika \textbf{6},
109 (1970) [Astrophys. \textbf{6}, 50 (1970)].

\end{enumerate}

\vspace{0.5cm}

\hfill \textit{Translated by V. Astakhov}

\end{document}